\documentclass[10pt,a4paper]{article}
 \pdfoutput=1
\usepackage{amsmath,fourier,amssymb,amsthm,graphicx,epstopdf,
color,cases}
\usepackage{chngcntr}
\counterwithout{figure}{section}
\usepackage[T1]{fontenc}
\allowdisplaybreaks
\numberwithin{equation}{section}

 \theoremstyle{plain}
 \newtheorem {hypo}{\bf\hspace{-\parindent}Hypothesis}[section]

 \newtheorem {prop}[hypo]{Proposition}
 
 \newtheorem {theo}[hypo]{Theorem}

 \newtheorem {conj}[hypo]{Conjecture} 
 
 \theoremstyle{remark}
 \newtheorem {rmk}[hypo]{Remark}
  
 \newcommand{\pf}{\begin{bpf}}

 \newcommand{\pfms}{\begin{bpfms}}
 \newcommand{\epf}{\end{bpf}\hfill$\square$\vspace{0.1cm}}
 \newcommand{\epfms}{\end{bpfms}\hfill$\square$\\ }
 \newcommand\ben{\begin{equation*}}
 \newcommand\ebn{\end{equation*}}
 \newcommand\beq{\begin{equation}}
 \newcommand\eeq{\end{equation}}
 \newcommand\ds{\displaystyle}
  \newcommand\lb{\left(}
  \newcommand\rb{\right)} 
  
   \newcommand\Cb{\mathbb{C}} 
   \newcommand\Zb{\mathbb{Z}}
   \newcommand\Pb{\mathbb{P}}

\usepackage{hyperref}

\begin{document}
\LARGE
\noindent
\textbf{On the connection problem for Painlev\'e I}
\normalsize
 \vspace{1cm}\\
 \textit{ 
 O. Lisovyy$\,^{a}$\footnote{lisovyi@lmpt.univ-tours.fr}, 
 J. Roussillon$\,^{a}$\footnote{julien.roussillon@lmpt.univ-tours.fr}}
 \vspace{0.2cm}\\
 $^a$  Laboratoire de Math\'ematiques et Physique Th\'eorique CNRS/UMR 7350,  Universit\'e de Tours, Parc de Grandmont,
  37200 Tours, France

  \begin{abstract}
  We study the dependence of the tau function of Painlevé I equation on the generalized monodromy of the associated linear problem. In particular, we compute connection constants relating the tau function asymptotics on five canonical rays at infinity. The result is expressed in terms of dilogarithms of cluster type coordinates on the space of Stokes data. 
  \end{abstract}
  
 \section{Introduction} 
 The present note is concerned with the first Painlevé equation, whose standard form reads
  \beq
  \label{PI}
  q_{tt}=6q^2+t.
  \eeq
 This equation represents the shortest entry of the Painlevé-Gambier list of 2nd order ODEs with the property that their solutions have no movable branch points. As is well-known, it appears as a similarity reduction of integrable PDEs such as KdV and Boussinesq equations, and also in the context of matrix models and two-dimensional quantum gravity, see \cite{FIKN,Kapaev3} for details and further references. Among more recent applications, let us mention that specific Painlevé I transcendents arise in the description of the universal behavior of solutions of the nonlinear Schr\"odinger equation \cite{DGK}, analysis of the cubic anharmonic oscillator \cite{Masoero}, and as a model for topological recursion \cite{IS}. General solution of Painlevé I has been conjecturally related to partition function of superconformal Argyres-Douglas theory of type $H_0$ \cite{BLMST}.  
 
  Painlevé I can be rewritten as a non-autonomous hamiltonian system 
  $q_t=H_p$, $p_t=-H_q$, where the time-dependent Hamiltonian is given by
  \beq\label{hamPI}
  H=\frac{p^2}{2}-2q^3-tq.
  \eeq
  The Hamiltonian itself satisfies the so-called $\sigma$-form of Painlevé I equation
  \beq \label{sigmaPI}
  H_{tt}^2=2\left(H-tH_t\right)-4 H_t^3,\eeq
  which is easily deduced by taking into account that $H_t=-q$. The Painlevé I tau function is defined
   up to a factor independent of $t$ 
  by
  \beq\label{tauPIdef}
  \partial_t\ln\tau=H.
  \eeq  
  
\begin{figure}[h!]
              \centering
              \includegraphics[height=3.5cm]{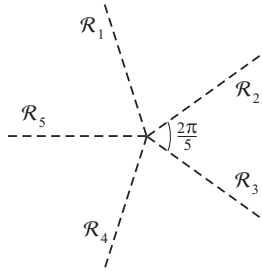}
              \caption{\label{CRsPI}
              Painlevé I canonical rays in the $t$-plane.}
              \end{figure}

  In the case of equation (\ref{PI}), the Painlevé property  means that
  every solution $q\lb t\rb$ is a meromorphic function  on the whole complex $t$-plane with only double poles, see e.g. \cite[Chapter 2]{Del} for a detailed proof. Moreover, the tau function $\tau\lb t\rb$ is holomorphic with only simple zeros. The asymptotic behavior of Painlevé I transcendents as $t\to \infty$ is rather intricate. The asymptotics is trigonometric along the canonical rays 
  $\mathcal R_k=\left\{\arg t=\pi-\frac{2\pi k}{5}\right\}$, $k\in\mathbb Z/5\mathbb Z$, whereas its generic form inside the sectors shown in Fig.~\ref{CRsPI} is described by the modulated elliptic functions \cite{Boutroux,JK}.

  The relation of Painlevé I to the theory of monodromy preserving de\-for\-mations \cite{JMU,FIKN} provides an avenue for solving the connection problem between different asymptotic directions at infinity. For that one needs to express the parameters of the asymptotic behavior in terms of the Stokes data of the associated linear problem. In the case of $q\lb t\rb$, the latter task has been accomplished by Kapaev \cite{Kapaev88,Kapaev93} on the canonical rays and by Kapaev and Kitaev \cite{KK} for the elliptic asymptotics. As far as the tau function is concerned, there remains a problem of evaluating certain constant factors. To explain what we have in mind, let us now formulate a sharp question concerning such {\it connection constants} in the asymptotics of $\tau\lb t\rb$. 
  
  As will be discussed below, the results of \cite{Kapaev88,Kapaev93} imply that for generic monodromy the asymptotic behavior of $\tau\lb t\rb$ on five canonical rays is given by
  \beq\label{tau5rays}
  \tau\lb t\to\infty\rb\simeq C_k x^{-\frac1{60}-\frac{\nu_k^2}{2}}e^{\frac{x^2}{45}+\frac{4}{5}i\nu_kx}
  \left[1+o\lb 1\rb\right],\qquad t\in\mathcal R_k,
  \eeq
  where $x=24^{\frac14}|t|^{\frac54}$. The parameters $\nu_1,\ldots,\nu_5$ may be expressed in terms of Stokes multipliers. Any pair of them can be taken as Painlevé I integrals of motion. Its knowledge fixes the other three $\nu_k$ as well as all subleading corrections to the asymptotics. The factors $C_k$ in (\ref{tau5rays}) are individually undefined since the equation (\ref{tauPIdef}) fixes $\tau\lb t\rb$ only up to a multiplicative constant. The ratios $C_k/C_{k'}$ describing relative tau func\-tion normalization on different rays are on the other hand unambiguously fixed by the appropriate Painlevé I function $q\lb t\rb$. Our main result, formulated in Theorem~\ref{mainth}, provides an explicit evaluation
  of these ratios in terms of asymptotic parameters $\nu_1,\ldots,\nu_5$.
  
  The computation of connection constants in the asymptotics of particular Painlevé tau functions was initially motivated by applications in random matrix theory and integrable systems, see e.g. \cite{BT,Tracy,Ehrhardt,DIK,DKV}
  and other references therein and in \cite{ILP}. Their systematic study has been initiated in \cite{ILT13,ILT14} where evaluations of these constants in terms of monodromy
  were conjectured for general solutions of Painlevé VI and Painlevé~III$_{D_8}$. The latter PIII$_{D_8}$ result has been proved by Its and Prokhorov \cite{IP} using the idea, first suggested in \cite{Bertola},
  of extending the Jimbo-Miwa-Ueno differential \cite{JMU} defining the isomonodromic tau function, to the space of monodromy data. General  construction of the localized formulas for this extended differential has been developed in \cite{ILP} and used there to derive connection constants for generic Painlevé VI and zero-parameter Painlevé II tau functions. The present work implements the approach outlined in \cite{ILP} in the case of Painlevé I.  
  \\ 

   \noindent
   { \small \textbf{Acknowledgements}.
   The present work  was supported by the PHC Sakura project 36175WA ``Isomonodromic deformations and conformal field theory''.
    We thank the organizers of the workshop ``Conformal Field Theory, isomonodromic tau-functions and Painlevé equations'' (Kobe, november 2016), where these results have been presented. O.L. would also like to thank Alexander Its for discussions and Andrei Kapaev for useful comments and help with the literature, in particular, for bringing \cite{Kapaev93} to authors' attention and informing them about a typo in \cite{Kapaev88}.}

 \section{Monodromy and quasiperiodicity}

 \subsection{Associated linear problem}
 Consider the system
 of linear differential equations
 \begin{subequations}
 \label{lpair1}
 \begin{numcases}{}
 \partial_z\Phi= A\lb z,t\rb\Phi,\label{lpair1a}\\
 \partial_{t\,}\Phi= B\lb z,t\rb\Phi,\label{lpair1b}
 \end{numcases}
 \end{subequations}
 with $A$ and $B$ given by
 \begin{subequations}
 \label{lpair1AB}
 \begin{align}
 \label{lpair1ABa}
 A\lb z,t\rb=&\,\lb\begin{array}{cc} 0 & 1 \\ 0 & 0 \end{array} \rb z^2+
    \lb\begin{array}{cc} 0 & q \\ 4 & 0 \end{array} \rb z
    +\lb\begin{array}{cc} -p & q^2+t/2 \\ -4q & p \end{array} \rb,\\
   B\lb z,t\rb=&\,\lb\begin{array}{cc} 0 & 1/2 \\ 0 & 0 \end{array} \rb z + \lb\begin{array}{cc} 0 & q \\ 2 & 0 \end{array} \rb.   
 \end{align}
 \end{subequations}
   Equations (\ref{lpair1a}), (\ref{lpair1ABa}) provide a canonical form
   for $2\times 2$ linear systems with a single irregular singular point  of Poincaré rank 3 on the Riemann sphere $\bar\Cb=\Pb^1\lb \Cb\rb$ with non-diagonalizable highest polar part; here the singularity is located at $\infty$. Parameters $p,q,t$ may be thought of as coordinates on the moduli space of appropriate irregular connections.
   The global asymptotic behavior of the fundamental matrix solution $\Phi\lb z,t\rb$ as ${z\to\infty}$ is characterized by a set of Stokes matrices which will be defined below. They constitute the generalized monodromy data for the linear system (\ref{lpair1a}). Requiring their invariance under simultaneous variation of 
  $t$, $p$ and $q$ leads to the second equation in the Lax pair~(\ref{lpair1}).
 The flatness condition 
  $\partial_t A-\partial_z B+\left[A,B\right]=0$ is equivalent to  Painlevé~I hamiltonian system described above.
 
  Another Lax pair frequently used in the literature is obtained from (\ref{lpair1AB}) by a combination of a gauge transformation and quadratic change of variable. Introducing
  \ben
  \Psi\lb\xi,t\rb=K\lb\xi\rb^{-1}\Phi\lb\xi^2,t\rb,\qquad
  K\lb\xi\rb=\lb\begin{array}{cc}
  \sqrt\xi/2 & \sqrt\xi/2 \\
  1/\sqrt\xi & - 1/\sqrt\xi
  \end{array}\rb,
  \ebn
  the new fundamental solution $\Psi\lb\xi,t\rb$ satisfies
   \begin{subequations}
   \label{lpair2}
   \begin{numcases}{}
   \partial_\xi\Psi= \tilde A\lb \xi,t\rb\Psi,\label{lpair2a}\\
   \partial_{t\,}\Psi= \tilde B\lb \xi,t\rb\,\Psi,\label{lpair2b}
   \end{numcases}
   \end{subequations}
   where the transformed matrices 
   \ben
   \tilde A\lb \xi,t\rb=2\xi K\lb\xi\rb^{-1}A\lb\xi^2,t\rb K\lb\xi\rb-K\lb\xi\rb^{-1}K'\lb\xi\rb,\qquad
   \tilde B\lb \xi,t\rb=K\lb\xi\rb^{-1}B\lb\xi^2,t\rb K\lb\xi\rb
   \ebn are explicitly given by
    \begin{subequations}
     \label{lpair2ABbis}
      \begin{align}
      \label{lpair2ABa}
     \tilde A\lb \xi,t\rb=&\,\lb 4\xi^4+2q^2+t\rb\sigma_z 
     -\lb2p \xi+\frac1{2\xi}\rb\sigma_x-\lb4q\xi^2+2q^2+t\rb i\sigma_y,\\
     \tilde B\lb \xi,t\rb=&\,\lb\xi+\frac q\xi\rb \sigma_z-\frac{iq\sigma_y}{\xi},
     \end{align}
    \end{subequations}
  and $\sigma_{x,y,z}$ denote the Pauli matrices,  
  \ben
  \sigma_x=\lb\begin{array}{cc}
  0 & 1 \\ 1 & 0
  \end{array}\rb,\qquad   \sigma_y=\lb\begin{array}{cc}
    0 & -i \\ i & 0
    \end{array}\rb,\qquad   \sigma_z=\lb\begin{array}{cc}
      1 & 0 \\ 0 & -1
      \end{array}\rb.
  \ebn
  The coeficient of the highest polar part of $\tilde A\lb \xi,t\rb$ at $\xi=\infty$ is diagonalizable (in fact, diagonal), which means that the corresponding irregular singular point is unramified.  This is the main advantage of the Lax pair (\ref{lpair2ABbis}) in comparison with (\ref{lpair1AB}).
  
  \subsection{Stokes data}
  The monodromy data of the linear system (\ref{lpair1a}) or its transformed version (\ref{lpair2a}) are integrals of motion of the Painlevé I equation which uniquely determine the solution $q\lb t\rb$ \cite{KK}. Let us now describe them in more detail.
  In the neighborhood of $\infty$, the equation (\ref{lpair2a}) possesses a unique formal solution
  of the form
  \begin{subequations}
  \begin{gather}  
  \Psi_{\mathrm{form}}\lb\xi \rb=G\lb\xi \rb e^{\Theta\lb \xi\rb},\\
  \label{Gxi}
  \Theta\lb \xi\rb=\lb\frac45\xi^5+t\xi\rb\sigma_z,\qquad 
  G\lb\xi \rb=\mathbb 1+\sum_{k=1}^{\infty}
    g_k\xi^{-k}.
  \end{gather}
  \end{subequations}
  The coefficients $g_k$ can in principle be iteratively determined from (\ref{lpair2a}). Below we will need the first 5 of these coefficients, explicitly given by
  \beq\label{gks}
  \begin{gathered}
  g_1=-H\sigma_z,\qquad
  g_2=\frac{H^2}{2}\mathbb 1+\frac q 2
  \sigma_x,\\
  g_3=\lb-\frac{H^3}{6}-\frac{2p-t^2}{24}\rb\sigma_z
  +\lb\frac{qH}{2}+\frac p4\rb i\sigma_y,\\
  g_4=\lb\frac{H^4}{24}+\frac{2p-t^2}{24} H
  +\frac{q^2}{8}\rb\mathbb 1 +\lb\frac{qH^2}{4}+\frac{pH}{4}+\frac{2q^2+t}{8}\rb\sigma_x,\\
  g_5=\lb-\frac{H^5}{120}-
  \frac{2p-t^2}{48}H^2-
  \frac{5q^2-2t}{40}H-\frac{4pq+1}{160}\rb\sigma_z+
  \lb\frac{qH^3}{12}+\frac{pH^2}{8}+\frac{2q^2+t}{8}H
  +\frac{2p-t^2}{48}q+\frac1{16}
  \rb i\sigma_y,
  \end{gathered} 
  \eeq
  where $H$ is the hamiltonian (\ref{hamPI}). 
  
          \begin{figure}[h!]
            \centering
            \includegraphics[height=6cm]{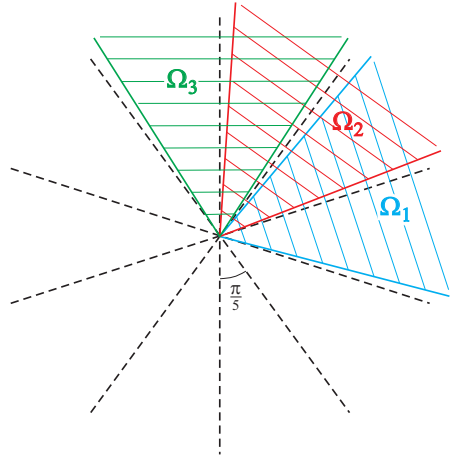}
            \caption{\label{sRays}
            Stokes sectors at $\infty$ in the $\xi$-plane.}
            \end{figure}
  
  Ten genuine canonical solutions $\Psi_{k}\lb\xi \rb$ are  uniquely specified by their
  asymptotic behavior 
  $\Psi_k\lb \xi\rb\simeq \Psi_{\mathrm{form}}\lb\xi \rb$ as $\xi\to\infty$  inside the Stokes sectors
  \ben
  \Omega_k=\left\{\xi\in\Cb\colon \frac{\lb 2k-3\rb\pi}{10}<\arg\xi< \frac{\lb 2k+1\rb\pi}{10}\right\},\qquad 
  k\in \Zb/10\Zb,
  \ebn
  see Fig.~\ref{sRays}.
  Canonical solutions are related by Stokes matrices, $S_k:={\Psi_{k}\lb\xi\rb}^{-1}\Psi_{k+1}\lb\xi\rb$. They are independent of $\xi$ and have a familiar triangular structure:
  \ben
  S_{2k-1}=\lb \begin{array}{cc}
      1 & s_{2k-1} \\ 0 & 1\end{array}\rb,\qquad
      S_{2k}=\lb \begin{array}{cc}
  1 & 0 \\ s_{2k} & 1\end{array}\rb,\qquad 
     k=1,\ldots ,5.
  \ebn
  
  Stokes parameters $s_{1\ldots 10}$ are not all independent. The non-transformed system (\ref{lpair1a}) has no singularity at $z=0$, which implies that
  \beq\label{parity}\Psi_k\lb\xi e^{i\pi}\rb=-i\sigma_x\Psi_k\lb\xi\rb.
  \eeq
  This in turn gives 
  \beq\label{per5}
  \sigma_x\Psi_{k+5}\lb\xi e^{i\pi}\rb\sigma_x=
  \Psi_k\lb \xi\rb. 
  \eeq
  Indeed, from (\ref{parity}) it follows that both sides of (\ref{per5}) satisfy the same equation (\ref{lpair1a}), hence to show their equality it suffices to compare their asymptotics inside the sector $\Omega_k$. Relation (\ref{per5}) implies that 
  $S_{k+5}=\sigma_x S_k\sigma_x$ and $s_{k+5}=s_k$. Furthermore, combining (\ref{parity}) and (\ref{per5}), one obtains a cyclic identity
  \beq\label{cyclic}
  S_1S_2S_3S_4S_5=i\sigma_x.
  \eeq
  It is equivalent to the equations 
   \beq\label{SDsym}
   s_{k+3}=i\lb 1+ s_k s_{k+1}\rb,\qquad s_{k+5}=s_k,
   \eeq
  which imply that there are at most 2 independent Stokes parameters. For example, for $s_2 s_3\ne -1$ one may express $s_1$, $s_4$, $s_5$ in terms of $s_2$, $s_3$:
    \beq\label{modpi}
    s_1=\frac{i-s_3}{1+s_2s_3},\qquad 
    s_4=\frac{i-s_2}{1+s_2s_3},\qquad
    s_5=i\lb1+s_2s_3\rb.
    \eeq
  The space $\mathcal S$ of Stokes data is therefore generically 2-dimensional\footnote{If the condition $s_2s_3\ne -1$ does not hold, then $s_2=s_3=i$, $s_5=0$, $s_1+s_4=i$, which corresponds to a 1-dimensional stratum of $\mathcal S$.}: these data can be expressed in terms of a pair of complex monodromy parameters which provide Painlevé I conserved quantities.   In what follows, we assume the genericity condition $s_k\ne 0$ for $k=1,\ldots ,5$, which excludes from our consideration the so-called {\it tronquées} solutions.

  Define  $v_k=-is_{2k}$ and further rewrite the relations (\ref{SDsym}) as
  \beq\label{SDsym2}
  v_{k-1}v_{k+1}=1-v_k.
  \eeq
  It is easy to check that the sequence defined by the latter equation  is 5-periodic. In fact, the recurrence (\ref{SDsym2}) describes mutations in the simplest rank $2$ cluster algebra of  finite type, associated to the Dynkin diagram~$A_2$. Also, introduce new monodromy parameters $\nu_k$ by
  \beq\label{defvk}
  v_k=e^{2\pi i \nu_k},\qquad k\in\Zb/5\Zb.
  \eeq
  It will turn out later that the most convenient sets of local coordinates on $\mathcal S$ are provided by the pairs $\lb \nu_k,\nu_{k+1}\rb$ associated to different clusters.

  Painlevé I transcendents can therefore be labeled as $q\lb t\,|\,\boldsymbol \nu\rb$, where $\boldsymbol \nu\in\mathcal S$ denotes the appropriate point in the space of Stokes data. The reader with no prior knowledge of Painlevé theory should think of $\boldsymbol \nu$ as of nonlinear analog of parameter $\alpha$
  in Bessel function $J_{\alpha}\lb t\rb$ or parameters $a,b,c$ in the hypergeometric function $_2F_1\lb a,b;c;t\rb$. It is important to recall that the tau function $\tau\lb t\,|\,\boldsymbol \nu\rb$ is so far defined only up to a multiplicative factor $\mathcal C\lb\boldsymbol \nu\rb$ independent of time.

 \subsection{$\Zb_5$-symmetry\label{secZ5}}
 In contrast to all other Painlevé equations, Painlevé I does not contain parameters and does not possess affine B\"acklund transformations. However it does have a finite $\Zb_5$-symmetry.
 If $q\lb t\rb$ and $H\lb t\rb$ are solutions of (\ref{PI}) and  (\ref{sigmaPI}), then clearly so are 
 $\tilde q\lb t\rb=\zeta^2 q\lb \zeta t\rb$ and $\tilde H\lb t\rb=\zeta H\lb \zeta t\rb$, where $\zeta=\exp\lb-\frac{2\pi i }5\rb$ 
 is a 5th root of unity. This in turn implies that if $\tau\lb t\rb$ is a Painlevé I tau function, then so is $\tilde\tau\lb t\rb=\tau\lb \zeta t\rb$.
 
 This symmetry can be lifted to solutions of the linear system (\ref{lpair2a})
 as $\Psi\lb \zeta \xi,t;\tilde q\lb t\rb\rb=
 \Psi\lb \xi,\zeta t; q\lb \zeta t\rb\rb$. As a consequence, Stokes parameters $\tilde s_k$, $\tilde v_k$ and $\tilde\nu_k$ corresponding to the transformed solution $\tilde q\lb t\rb$ are expressed as $\tilde s_k=s_{k+2}$, $\tilde v_k=v_{k+1}$ and $\tilde\nu_k=\nu_{k+1}$. Introducing the operator $T$ of cyclic permutation which acts on monodromy parameters as $T\nu_k=\nu_{k+1}$ with $k=1,\ldots,5$, we may then write
 \beq\label{backlundPI}
  q\lb \zeta t \,|\,\boldsymbol\nu\rb=\zeta^3q\lb  t\,|\, T\boldsymbol\nu\rb.
 \eeq
 The analog of this relation for the tau function is
 \beq\label{ccdef}
 \tau\lb\zeta t\,|\,\boldsymbol\nu\rb=\Upsilon\lb\boldsymbol \nu \rb\cdot 
 \tau\lb t\,|\,T\boldsymbol\nu\rb.
 \eeq
 The appearance of the prefactor $\Upsilon\lb\boldsymbol\nu\rb$ is related to the ambiguity in the definition of the tau function by solution of the $\sigma$-Painlevé I equation (\ref{sigmaPI}). However, once the normalization of the tau function $\tau\lb t\,|\,\boldsymbol\nu\rb$ is fixed, the connection coefficient $\Upsilon\lb\boldsymbol\nu\rb$ becomes a well-defined function of monodromy. 
 
 One way to choose the normalization is to require $\tau\lb 0\,|\,\boldsymbol\nu\rb=1$, so that we trivially have  $\Upsilon\lb\boldsymbol\nu\rb=1$. Although this is legitimate in the generic situation where $t=0$ is not a pole of $q\lb t\,|\,\boldsymbol\nu\rb$, it is more natural, both conceptually and from the point of view of applications, to normalize the asymptotic behavior of the tau function at the only genuine  Painlevé I singular point  $t=\infty$. 
  Our main goal in the next subsections is the determination of the explicit form of $\Upsilon\lb\boldsymbol\nu\rb$ in this setting.

 \section{Extended Painlevé I tau function}

 \subsection{General setup}

 A normalization of the Painlevé I tau function can be introduced by constructing a closed 1-form $\hat\omega\in\Lambda^1\lb \Cb\times\mathcal S\rb$ whose restriction to the first factor coincides with $H dt$. Then the tau function is defined up to a constant independent of monodromy data by
 \ben
 \tau\lb t\,|\,\boldsymbol\nu\rb=\exp\int \hat\omega.
 \ebn
 A general approach for constructing $\hat\omega$ has been developed in \cite{ILP} using earlier results of the works \cite{Bertola,IP}. It can be summarized as follows:
 \begin{itemize}
 \item For a linear system $\partial_{\xi}\Phi=A\lb \xi\rb\Phi$ with rational 
 $A\lb \xi\rb$, one should write formal solutions at each singular point $a_{i}$ of $A\lb \xi\rb$  as
 $\Phi^{(i)}_{\text{form}}\lb \xi\rb= G^{(i)}\left(\xi\right)
 e^{\Theta_{i}\lb \xi\right)}$, where
 \begin{subequations}\label{formGTH}
 \begin{align}
  G^{(i)}\left(\xi\right)=&\,G_{i}\left[\mathbb 1+\sum_{k=1}^{\infty}g_{i,k}\left(\xi-a_{i}\right)^k\right],\\
  \Theta_{i}\lb \xi\right)=&\,\sum_{k=-r_{i}}^{-1}
  \Theta_{i,k}\frac{\left(
  \xi-a_{i}\right)^k}{k}+\Theta_{i,0}\ln\left( \xi-a_{i}\rb.
 \end{align}
 \end{subequations}
 In the last formula, $r_{i}$ denotes the Poincaré rank of $a_{i}$ and the matrices $\Theta_{i,k}$ are all diagonal. Their elements (with the exception of  $\Theta_{i,0}$) and positions $a_{i}$ play the role of isomonodromic times. For a singular point at~$\infty$, the expressions (\ref{formGTH}) should be appropriately modified.
 \item Define a $1$-form $\omega$ by
 \beq\label{extomega}
 \omega=\sum_{i}\operatorname{res}_{\xi=a_{i}}\operatorname{Tr}
 \lb A\lb\xi\right) dG^{\lb i\right)}\left(\xi\right){G^{\lb i\right)}\left(\xi\right)}^{-1}\right).
 \eeq
 where the differential $d=d_{\mathcal T}+d_{\mathcal M}$ is taken both with respect to times $\mathcal T$ \textit{and}
 monodromy parameters~$\mathcal M$. The differential $\Omega=d\omega$ was shown in \cite{ILP} to be a closed $2$-form on $\mathcal M$  only, which is furthermore independent of isomonodromic times.
 \item The $2$-form $\Omega$ can in principle be calculated explicitly using the asymptotics of solutions of the deformation equations expressed in terms of monodromy. Once the expression for $\Omega$ is found, one may look for a $1$-form $\omega_0$ on $\mathcal M$ such that $d\omega_0=\Omega$ and define $\hat\omega=\omega-\omega_0$. Of course, $\omega_0$ is determined by $\Omega$ up to addition of an exact differential on $\mathcal M$, which is the origin of the tau function normalization ambiguity.
 \end{itemize}
 
 The results of \cite{ILP} have been obtained under assumption that the  highest polar contribution to $A\lb \xi\rb$ at each singular point is diagonalizable; for Fuchsian singularities, it is required in addition to be  non-resonant. The matrix $A\lb z,t\rb$ given by (\ref{lpair1ABa})  violates the diagonalizability condition at $z=\infty$, whereas its transformed version $\bar A\lb\xi,t\rb$ from
 (\ref{lpair2ABa}) has a resonant (though trivial) Fuchsian singularity at $\xi=0$.
 
  We are not going to develop a general theory for the ramified irregular singularities in this paper. Instead, consider the formula (\ref{extomega}) as an ansatz for $\omega$
  for the transformed system (\ref{lpair2a}), simply ignoring the undefined contribution of the resonant singular point $\xi=0$. We thus introduce
  \ben
  \omega:=\operatorname{res}_{\xi=\infty} \operatorname{Tr}\lb
  \bar A\lb \xi,t\rb dG\lb\xi\rb {G\lb\xi\rb}^{-1}\rb,
  \ebn
  with $G\lb \xi\rb$ defined by (\ref{Gxi}). Since $\bar A\lb\xi,t\rb$ is a Laurent polynomial of degree $4$ in $\xi$, the residue is given by
  \beq\label{omega01}
  \begin{aligned}
  \omega=\operatorname{Tr}\biggl\{& -4\sigma_z\lb dg_1 h_4+dg_2 h_3+dg_3h_2+dg_4h_1+dg_5\rb
  +4iq\sigma_y\lb dg_1 h_2+dg_2 h_1+dg_3\rb+\biggr.\\
  &\biggl.\,+\,2p\sigma_x\lb
  dg_1 h_1+dg_2\rb-\lb 2q^2+t\rb\lb \sigma_z-i\sigma_y\rb dg_1\biggr\},
  \end{aligned}
  \eeq
  where  $h_k$ denote the expansion coefficients of the inverse matrix
  ${G\lb\xi\rb}^{-1}=\mathbb 1+\sum_{k=1}^{\infty}h_k\xi^{-k}$.   Let us express them in terms of $g_k$, 
  \begin{gather*}
  h_1=-g_1,\qquad h_2=-g_2+g_1^2,\qquad h_3=-g_3+g_2g_1+g_1g_2-g_1^3,\\
  h_4=-g_4+g_3g_1+g_1g_3+g_2^2-g_2g_1^2-g_1g_2g_1-g_1^2g_2+g_1^4,
  \end{gather*} 
  and substitute into (\ref{omega01}). Using explicit expressions (\ref{gks}) for $g_1,\ldots, g_5$, after a lengthy but straightforward simplification we find that
  \begin{gather}\label{omega02}
  \omega=2\lb H dt+Q_adm_a+Q_b dm_b\rb,
  \end{gather}
  where $m_{a,b}$ are two arbitrary independent local coordinates
  on the space $\mathcal S$ of Painlevé I Stokes data, and the coefficients $Q_{a,b}$ are given by
  \begin{align}
  \label{omega03}
  \nonumber Q_k=&\,\frac15\lb 4tq_t q_{tm_k}+
  3q_tq_{m_k}-2qq_{tm_k}-24tq^2q_{m_k}-4t^2q_{m_k}\rb=\\
  =&\,\frac15\lb 4t H_{m_k}+
    3q_tq_{m_k}-2qq_{tm_k}\rb, \qquad\qquad\qquad k=a,b.
  \end{align}
  It can now be checked directly that  $\omega$ indeed has all the necessary properties for realization of the scheme outlined above.
  Note, however, the appearance of an extra factor of $2$ in (\ref{omega02}).
  \begin{prop}\label{prom}
  The differential $\Omega=d\omega$ of the form $\omega$ defined by (\ref{omega02})--(\ref{omega03}) is a closed constant $2$-form on 
  $\mathcal S$.
  \end{prop}
  \pf Differentiating $H$ with respect to monodromy, one obtains
  \ben
  \partial_{m_k}H=q_tq_{tm_k}-\lb 6q^2+t\rb q_{m_k}.
  \ebn
  Compute the time derivative of (\ref{omega03}) and use Painlevé I equation (\ref{PI}) to eliminate all 2nd order time derivatives in the resulting expressions. Simplifying the result, we observe that $\partial_tQ_k=\partial_{m_k}H$,  which shows that $\Omega$ is a 2-form on $\mathcal S$ only. 
  
   One also has
  \begin{align*}
  \partial_{m_a}Q_b=&\,
  \frac15\,\Bigl( 4tq_{tm_a} q_{tm_b}+4tq_t q_{tm_am_b}+
    3q_tq_{m_am_b}+3q_{tm_a}q_{m_b}\Bigr.\\
    & \Bigl.-
    2qq_{tm_a m_b}-
        2q_{m_a}q_{tm_b}-24tq^2q_{m_a m_b}-48tq q_{m_a}q_{m_b}-4t^2q_{m_am_b}\Bigr).
  \end{align*}  
  Almost all terms in this formula are symmetric with respect to the exchange $a\leftrightarrow b$ and therefore do not contribute to $\Omega=2\lb \partial_{m_a}Q_b-\partial_{m_b}Q_a\rb dm_a\wedge dm_b$. The non-symmetric part yields
  \beq\label{bigomega}
  \Omega=2\lb q_{tm_a}q_{m_b}-q_{tm_b}q_{m_a}\rb
   dm_a\wedge dm_b.
  \eeq
  It remains to show that $\partial_t\lb q_{tm_a}q_{m_b}-q_{tm_b}q_{m_a}\rb=0$, which is an easy consequence
  of (\ref{PI}).
  \epf
   
 \subsection{Asymptotics on 5 canonical rays}
   Our task in this subsection is to construct a 1-form
   $\omega_0\in\Lambda^1\lb\mathcal S\rb$ such that $d\omega_0=\Omega$.
   In order to achieve this goal, we will first compute the explicit form of $\Omega$ in terms of Stokes parameters using the results of Kapaev \cite{Kapaev88,Kapaev93} (see also \cite{Takei}) for the asymptotics of Painlevé I transcendents on the rays 
   \ben
   \mathcal R_k=\left\{t\in e^{\pi i -\frac{2\pi i k}{5}}\mathbb R_{\ge0}\right\},\qquad k\in\Zb/5\Zb,
   \ebn 
   as $|t|\to\infty$. Recall that the Stokes multipliers are parameterized as 
   \beq\label{SDparam}
   s_{2k}=ie^{2\pi i \nu_k},\qquad k\in\Zb/5\Zb.
   \eeq
   The pairs $\lb\nu_k,\nu_{k+1}\rb$ correspond to 5 different choices of local coordinates on $\mathcal S$, adapted for description of the asymptotics on different rays $\mathcal R_k$.
   
   \begin{theo}[\cite{Kapaev93}]
   The asymptotic behavior of Painlevé I function $q\lb t\,|\,\boldsymbol\nu\rb$ as
   $|t|\to\infty$, $\arg t=\pi-\frac{2\pi k}{5}$ with $ k\in\Zb/ 5\Zb$ is described by the following formulas:
   \begin{itemize}
   \item For $\left|\Re\nu_k\right|<\frac16$, one has
   \beq\label{as01}
   q\lb t\,|\,\boldsymbol\nu\rb=
   e^{\frac{4\pi i k}{5}}\sqrt{ e^{\frac{2\pi i k}{5}-\pi i}t}\left[-\frac1{\sqrt6}+
   \sum_{\epsilon=\pm}\alpha_\epsilon\lb\nu_k\rb x^{-\frac{1+2\epsilon\nu_k}{2}}e^{\frac{4i\epsilon x}{5}+2\pi i\epsilon\nu_{k+1}}+
   O\lb x^{-1+2\left|\Re\nu_k\right|}\rb\right],
   \eeq
   where $x=24^{\frac14}\lb  e^{\frac{2\pi i k}{5}-\pi i}t\rb^{\frac54}$
   and the coefficients $\alpha_{\pm}\lb\nu\rb$ are given by
   \beq\label{as02}
   \alpha_+\lb\nu\rb=\frac{48^{-\nu}e^{-\frac{i\pi\lb1+2\nu\rb}{4}}
   \Gamma\lb 1+\nu\rb}{2\sqrt\pi},
   \qquad \alpha_-\lb \nu\rb=
   \frac{48^{\nu}e^{-\frac{i\pi\lb1-2\nu\rb}{4}}
     \sqrt\pi }{ \Gamma\lb \nu\rb}.
   \eeq
   \item For $0<\Re\nu_k<1$, one has
   \beq\label{as03}
    q\lb t\,|\,\boldsymbol\nu\rb=
      e^{\frac{4\pi i k}{5}}\sqrt{ e^{\frac{2\pi i k}{5}-\pi i}t}\left[-\frac1{\sqrt6}+\frac{\sqrt6}{1 +\sum_{\epsilon=\pm}\beta_\epsilon\lb\nu_k\rb x^{\frac{(1-2\nu_k)\epsilon}{2}} e^{\frac{4i\epsilon x}{5}+2\pi i\epsilon\nu_{k+1}}
     +O\lb x^{2\left|\Re\nu_k\right|-2} \rb}\right],
   \eeq
   where $\beta_+\lb\nu\rb=\frac{\sqrt6}{\alpha_-\lb\nu\rb}$ and
   $\beta_-\lb\nu\rb=\frac{\alpha_-\lb\nu\rb}{4\sqrt6}$.
   \end{itemize}
   \end{theo}
   
    Of course, two behaviors (\ref{as01}) and (\ref{as03}) are compatible on the overlap of the corresponding domains. The latter cover all possible values of Stokes parameters except pathological one-dimensional strata described above. It turns out that both formulas produce the same asymptotic form of the tau function. Moreover, iteratively computing next terms in the expansion of $\tau\lb t\,|\,\boldsymbol\nu\rb$, one may observe the following periodic pattern\footnote{Let us mention the paper \cite{HRZ} which is concerned with another type of Painlevé I tau function expansions; namely, around movable poles of $q(t\,|\,\boldsymbol\nu)$, i.e. zeros of $\tau(t\,|\,\boldsymbol\nu)$. This study highlights the fact that Painlevé I can be considered as a deautonomization of the differential equation for the Weierstrass $\wp$-function.}:
    
  \begin{conj}[\cite{BLMST}, Section 3.1]\label{conjecPI}
  Asymptotic expansion of the Painlevé I tau function  $\tau\lb t\,|\,\boldsymbol\nu\rb$ as $t\to\infty$ along the ray $\mathcal R_k$ has the structure of a Fourier transform,
  \begin{subequations}
 \begin{align}
 \label{conj01}&\tau(t\,|\,\boldsymbol\nu)\simeq \mathcal C_k\lb\boldsymbol\nu\rb x^{-\frac{1}{60}}e^{\frac{x^2}{45}}\sum_{n\in\mathbb Z}e^{2\pi in\nu_{k+1}}
 \mathcal{B}\left(\nu_k+n,x\right),\\
 \label{conj02}&\mathcal{B}\left(\nu,x\right)\simeq C\left(\nu\right)x^{
 -\frac{\nu^2}{2}}
 e^{\frac{4}{5}\,i\nu x}\left[1+\sum_{k=1}^{\infty}\frac{
 B_k\left(\nu\right)}{x^k}\right],\\
 \label{conj03}& C\left(\nu\right)=48^{-\frac{\nu^2}{2}}
 \lb 2\pi \rb^{-\frac{\nu}{2}}e^{-\frac{i\pi \nu^2}{4}} G\left(1+\nu\right),
 \end{align}
 \end{subequations}
 where as before  $x=24^{\frac14}\lb  e^{\frac{2\pi i k}{5}-\pi i}t\rb^{\frac54}$, and $G\lb z\rb$ denotes the Barnes $G$-function.
  \end{conj}  
 This proposal has been verified by calculating explicitly over 50 first terms in the asymptotic expansion of $\tau\lb t\,|\,\boldsymbol\nu\rb$. Setting for definiteness  $\left|\Re\nu_k\right|\leq \frac12$, this corresponds to taking the values $|n|\leq 4$ in (\ref{conj01}) and going up to
 $k=7$ in (\ref{conj02}).
 The coefficients $B_k\lb\nu\rb$ of $\mathcal{B}\left(\nu,t\right)$ are polynomials of degree $3k$ in $i\nu$  with rational coefficients; the first few of them are given by
    \begin{align*}
    &B_1\left(\nu\right)=-\frac{i\nu\left(94\nu^2+17\right)}{96},\\
    &B_2\left(\nu\right)=-\frac{44\,180\nu^6+170\,320\nu^4+
    74\,985\nu^2+1\,344}{92\,160},\\
    &B_3\left(\nu\right)=\frac{i\nu\lb
     4\,152\,920 \nu^8+ 45\,777\,060 \nu^6
     + 156\,847\,302 \nu^4 + 124\,622\,833 \nu^2+
    13\,059\,000   \rb}{26\,542\,080},\\
    &\ldots\quad\ldots\quad\ldots
    \end{align*} 

 We are now in a position to determine the explicit form of $\Omega$ defined by (\ref{bigomega}). It is  very useful to rewrite the latter formula  as
 \beq\label{bigomegav2}
 \Omega=2d_{\mathcal S}q_t\wedge d_{\mathcal S}q,
 \eeq 
 where $d_{\mathcal S}$ denotes the differential taken with respect to the Stokes data. 
 Formal expansion of $q\lb t\,|\,\boldsymbol\nu\rb$ on the ray $\mathcal R_k$ can be written as
 \ben
 q\lb t\,|\,\boldsymbol\nu\rb=
    e^{\frac{4\pi i k}{5}}\sqrt{ e^{\frac{2\pi i k}{5}-\pi i}t}\left[-\frac1{\sqrt6}+\sum_{l=1}^{\infty}\sum_{j=0}^{l}
    \alpha_{l,l-2j}\frac{p^{l-2j}\lb x\rb}{x^{l/2}}\right],
 \ebn 
 where
 \ben
 p\lb x\rb = x^{-\nu_k}e^{\frac{4ix}{5}},\qquad 
 x=24^{\frac14}\lb  e^{\frac{2\pi i k}{5}-\pi i}t\rb^{\frac54}.
 \ebn
 Although the structure of this expansion is not as transparent as for the tau function $\tau\lb t\,|\,\boldsymbol\nu\rb$, we only need a few first terms of it. It suffices to compute $\Omega$ under assumption $|\Re\nu_k|<\frac16$, in which case the terms present in (\ref{as01}), i.e. with $l=1$, are already sufficient. Their straightforward substitution into (\ref{bigomegav2}) gives
 \beq\label{bigomegav3}
 \Omega=-4ix \, d_{\mathcal S}\lb \alpha_+\lb\nu_k\rb x^{-\frac{1+2\nu_k}{2}}e^{\frac{4ix}{5}+2\pi i\nu_{k+1}}\rb\wedge
 d_{\mathcal S}\lb
    \alpha_-\lb\nu_k\rb x^{-\frac{1-2\nu_k}{2}}e^{-\frac{4ix}{5}-2\pi i\nu_{k+1}}\rb +O\lb
    x^{-\frac12+3\left|\Re\nu_k\right|}\rb.
 \eeq
 Using that the coefficients $\alpha_{\pm}\lb\nu_k\rb$ are independent of $\nu_{k+1}$, the first term in (\ref{bigomegav3}) can be simplified to $-8\pi d\bigl(\alpha_+\lb\nu_k\rb
 \alpha_-\lb\nu_k\rb\bigr)\wedge d\nu_{k+1}$, and further reduced to  $4\pi i\,d\nu_k\wedge d\nu_{k+1}$ using (\ref{as02}). The error term is actually absent as it has been shown in Proposition~\ref{prom}
 that $\Omega$ is independent of $t$. We thus obtain
 \begin{prop}
 The 2-form $\Omega$ can be expressed as 
 \beq
 \label{bigomega04}
 \qquad\Omega=4\pi i\, d\nu_k\wedge d\nu_{k+1}, 
 \qquad\qquad  k\in\Zb/5\Zb,
 \eeq
 where $\lb\nu_k,\nu_{k+1}\rb$ is any of the 5 pairs of local coordinates on $\mathcal S$ defined by (\ref{SDparam}).
 \end{prop}
 In other words, each of the 5 pairs $\lb\nu_k,\nu_{k+1}\rb$ provides canonical coordinates on the space $\mathcal S$ of Stokes data of the linear system associated with Painlev\'e I. One may also rewrite the formula (\ref{bigomega04}) directly in terms of the Stokes parameters,
 \beq
 \qquad\Omega=\frac{i}{\pi}\frac{ds_{2+2k}\wedge ds_{3+2k} }{1+s_{2+2k}s_{3+2k}},\qquad\qquad
  k\in\Zb/5\Zb.
 \eeq
 
 The tau function normalization is determined up to a factor independent on monodromy by the choice of a form $\omega_0\in\Lambda^1\lb\mathcal S\rb$ such that
 $d\omega_0=\Omega$; recall that there is a freedom of adding to
 $\omega_0$ an exact differential on $\mathcal S$. In principle we could already set e.g. $\omega_0=4\pi i\nu_k d\nu_{k+1}$, and define the  extended tau function on $\Cb\times\mathcal S$ as $d\ln\tau=\lb\omega-\omega_0\rb/{2}$. This choice of $\omega_0$ turns out to be compatible with setting $\mathcal C_k\lb\boldsymbol\nu\rb=1$ in
 the asymptotics (\ref{conj01}) on the ray $\mathcal R_k=e^{\pi i-\frac{2\pi i k}{5}}\mathbb R_{\ge 0}$.
 
 \begin{prop}\label{propp}
 Given $k\in\Zb/5\Zb$, let us introduce
  \beq
  \omega_{0,k}:=4\pi i\nu_k d\nu_{k+1}.
  \eeq 
 Let $\left|\Re\nu_k\right|<\frac12$.
 The extended tau function $\tau_k\lb t\,|\,\boldsymbol\nu\rb$ 
 defined by
 \beq
 d\ln\tau_k\lb t\,|\,\boldsymbol\nu\rb=\frac{\omega-\omega_{0,k}}{2},
 \eeq
 with $\omega$ given by (\ref{omega02})--(\ref{omega03}),
 is characterized by the following asymptotic behavior as $t\to \infty$ along $\mathcal R_k$:
 \beq\label{normC}
 \tau_k\lb t\,|\,\boldsymbol\nu\rb=\mathcal C_k\cdot 
 C\lb\nu_k\rb x^{-\frac1{60}-\frac{\nu_k^2}2}e^{\frac{x^2}{45}
 +\frac45 i\nu_k x}\Bigl[1+o\lb 1\rb\Bigr],
 \eeq
 where $x=24^{\frac14}\lb e^{\frac{2\pi i k}{5}-\pi i}t\rb^{\frac54}$ and $\mathcal C_k$ is independent of Stokes data.
 \end{prop}
 \pf Considering the leading  terms in (\ref{conj01})--(\ref{conj02})
 (note that the conjectural part of the statement concerns only the  full expansion), one finds that, as $t\to e^{\pi i-\frac{2\pi i k}{5}}\infty$,
 \begin{align*}
 d\ln\tau_k\lb t\,|\,\boldsymbol\nu\rb=d\lb\frac{x^2}{45}+\frac{4i\nu_k x}5-\frac{\nu_k^2\ln x}{2}+\ln C\lb\nu_k\rb\rb+d\ln \mathcal C_k\lb \boldsymbol\nu\rb+o\lb 1\rb.
 \end{align*} 
 On the other hand, from (\ref{omega02})--(\ref{omega03}) and
 the asymptotics (\ref{as01}) one may deduce the corresponding asymptotics of~$\omega$. From the expressions (\ref{as02}) for $\alpha_{\pm}\lb \nu\rb$ combined with the classical formula 
 \beq\label{dBarnesG}
 d\ln G\lb1+ z\rb=zd\ln\Gamma\lb1+ z\rb-d\lb \frac{z^2}{2}\rb+
   \frac{\ln 2\pi-1}{2}dz,
 \eeq
 after somewhat lengthy simplification it follows  that
 \ben
 \omega=2d\lb\frac{x^2}{45}+\frac{4i\nu_k x}5-\frac{\nu_k^2\ln x}{2}+\ln C\lb\nu_k\rb\rb+4\pi i \nu_k d\nu_{k+1} +o\lb1\rb,
 \ebn
 which yields the statement of the proposition.
 \epf

 \subsection{Connection constant}
 Let us now set $\mathcal C_k=1$ in (\ref{normC}). This defines five distinct tau function normalizations $\tau_{k}\lb t\,|\,\boldsymbol\nu\rb$ corresponding to normalized asymptotic behaviors on different rays. The connection coefficients that we are after can be alternatively defined as
 \beq\label{Upskkp}
 \Upsilon_{kk'}\lb \boldsymbol\nu\rb=\frac{\tau_{k'}\lb t\,|\,\boldsymbol\nu\rb}{
 \tau_{k}\lb t\,|\,\boldsymbol\nu\rb}.
 \eeq
 Proposition~\ref{propp} implies that
 \ben
 d\ln\Upsilon_{kk'}\lb \boldsymbol\nu\rb=
 \frac{\omega_{0,k}-\omega_{0,k'}}{2}=
 2\pi i\lb\nu_k d\nu_{k+1}-\nu_{k'}d\nu_{k'+1}\rb.
 \ebn
 Thus $\ln\Upsilon_{kk'}\lb \boldsymbol\nu\rb$ coincides with the generating function of the canonical transformation between the pairs  $\lb \nu_k,\nu_{k+1}\rb$ and $\lb \nu_{k'},\nu_{k'+1}\rb$ of local Darboux coordinates
 on Painlevé I monodromy manifold. To obtain its explicit form (up to an additive constant independent of monodromy), it clearly suffices to compute the antiderivative 
 \beq\label{auxfo}
 2\pi i \int\lb\nu_{k-1} d\nu_{k}-\nu_{k}d\nu_{k+1}\rb,
 \eeq
 which enters into the expression for the connection constant
 $\Upsilon_{k-1,k}\lb \boldsymbol\nu\rb$ between adjacent rays. 
  
 We are now finally ready to state our main result:
 \begin{theo}\label{mainth} The connection coefficient $\Upsilon_{k-1,k}\lb \boldsymbol\nu\rb$ is expressed in terms of Stokes data as
 \beq\label{UpsPI}
 \Upsilon_{k-1,k}\lb \boldsymbol\nu\rb=e^{\frac{\pi i}{30}}
 \lb 2\pi\rb^{-\nu_{k}}e^{2\pi i \nu_{k-1}\nu_{k}-\frac{\pi i \nu_{k}^2}{2}}\hat{G}\lb\nu_{k}\rb,
 \eeq
 where $\hat G\lb z\rb=\ds\frac{G\lb 1+z\rb}{G\lb 1-z\rb}$ and
 $k\in\Zb/5\Zb$.
 \end{theo}
 \pf  Let us rewrite the recurrence relation (\ref{SDsym2}) in terms of $\nu_k$ with the help of (\ref{defvk}):
 \beq\label{recnuk}
 e^{2\pi i \nu_{k+1}}=\lb 1- e^{2\pi i \nu_{k}}\rb e^{-2\pi i\nu_{k-1}}.
 \eeq
 This transforms the antiderivative (\ref{auxfo}) into
 \ben
 \ln \Upsilon_{k-1,k}\lb \boldsymbol\nu\rb =
 2\pi i\nu_{k-1}\nu_{k}-\int \nu_{k}d\ln\lb 1-e^{2\pi i \nu_{k}}\rb.
 \ebn
 The identity (\ref{dBarnesG}) implies 
 the differentiation formula $d\ln\hat G\lb z\rb=\ln2\pi\,dz-zd\ln\sin\pi z$. Using it to compute the last integral, we obtain
 \ben
  \Upsilon_{k-1,k}\lb \boldsymbol\nu\rb=
  \chi\cdot\lb 2\pi\rb^{-\nu_{k}}e^{-\frac{\pi i \nu_{k}^2}{2}+2\pi i \nu_{k-1}\nu_{k}}\hat{G}\lb\nu_{k}\rb,
 \ebn
 where $\chi$ is a numerical constant independent of monodromy. Analogous constants for Painlevé VI and Pain\-levé~II have been fixed in \cite{ILP} with the help of special solutions (respectively, algebraic/Picard and Hastings-McLeod) of the corresponding equations. Even though such solutions are not available for Painlevé I, we will be able to find an explicit evaluation for $\chi$ by exploiting the $\Zb_5$-symmetry.
 
 Denote by $\tau^{(0)}\lb t\,|\,\boldsymbol\nu\rb$ the Painlevé I tau function associated to Stokes data $\boldsymbol\nu$ and
 normalized as $\tau^{(0)}\lb 0\,|\,\boldsymbol\nu\rb=1$. Recall that $\tau^{(0)}\lb\zeta t\,|\,\boldsymbol\nu\rb= 
  \tau^{(0)}\lb t\,|\,T\boldsymbol\nu\rb$, where $\zeta=\exp\lb-\frac{2\pi i}{5}\rb$ and $T$ cyclically permutes Stokes parameters, cf Sub\-section~\ref{secZ5}. We can then write
  \ben
  \tau_{k}\lb t\,|\,\boldsymbol\nu \rb=\tau_0\lb \zeta^{-k}t\,\bigl|\,T^k\boldsymbol\nu\rb=\tilde\Upsilon\bigl(
  T^k\boldsymbol\nu\bigr)\, \tau^{(0)}\lb \zeta^{-k}t\,\bigl|\,T^k\boldsymbol\nu\rb=
  \tilde\Upsilon\bigl(
    T^k\boldsymbol\nu\bigr)\,\tau^{(0)}\lb t\,\bigl|\,\boldsymbol\nu\rb,
  \ebn
  where $\tilde\Upsilon\lb\boldsymbol\nu\rb$ stands for the coefficient of relative normalization of the tau functions $\tau_0$ and $\tau^{(0)}$ (i.e. connection constant between $-\infty$ and $0$). As a consequence, the coefficients (\ref{Upskkp}) have the structure
  \ben
  \Upsilon_{kk'}\lb\boldsymbol\nu\rb=\frac{\tilde\Upsilon\bigl(
      T^{k'}\boldsymbol\nu\bigr)}{\tilde\Upsilon\bigl(
          T^k\,\boldsymbol\nu\bigr)}.
  \ebn
  Although explicit form of $\tilde\Upsilon\lb\boldsymbol\nu\rb$ is unknown, this structure implies that for a point $\boldsymbol\nu^{f}\in\mathcal S$ fixed by $T$ we should have $\Upsilon_{kk'}\lb\boldsymbol\nu^{f}\rb=1$. One has however to check that the tau function $\tau\lb t\,|\,\boldsymbol\nu^f\rb$ associated to this specific monodromy does not vanish at $t=0$ (or, equivalently, that $q\lb t\,|\,\boldsymbol\nu^f\rb$ does not have a pole there). This condition ensures the existence of $\tau^{(0)}$ which appears in the above argument, and it is verified for at least one of the two fixed points of $T$, namely, for
  \ben
  v_1=\ldots=v_5=e^{2\pi i\nu_f}=\frac{\sqrt5-1}{2}.
  \ebn
  It follows that the numerical constant we are looking for is given by
  \beq\label{chiau1}
  \chi=\lb 2\pi\rb^{\nu_f}e^{-\frac{3\pi i \nu_f^2}{2}}\hat G\lb -\nu_f\rb.
  \eeq
  
  In order to further simplify this representation, let us note that the function $\hat{G}\lb z\rb$ is closely related to the classical dilogarithm 
  $\operatorname{Li}_2\lb e^{2\pi i z}\rb$:
  \beq\label{chiau2}
  \hat G\lb z\rb=\lb\frac{\sin\pi z}{\pi}\rb^{-z}\exp\left\{\frac{\pi i z\lb 1-z\rb}{2}-\frac{\pi i}{12}-\frac{\operatorname{Li}_2\lb e^{2\pi i z}\rb}{2\pi i}\right\}.
  \eeq
  The relevant quantity 
  \beq\label{chiau3}
  \operatorname{Li}_2\lb\frac{\sqrt5-1}{2}\rb=\frac{\pi^2}{10}-
  \ln^2\frac{\sqrt5+1}{2}.
  \eeq
  is one of the few dilogarithm values with explicit elementary evaluations.
  Putting (\ref{chiau1})--(\ref{chiau3}) together and simplifying the result, we end up with  $\chi=e^{\frac{\pi i }{30}}$.
 \epf
 \begin{rmk}
 Using the recurrence relation $\hat G\lb z+1\rb=-\frac{\pi}{\sin\pi z}\hat G\lb z\rb$ and the formula (\ref{recnuk}), it is straightforward to check that the answer (\ref{UpsPI}) satisfies the quasiperiodicity relations
  \begin{align*}
  &\Upsilon_{k-1,k}\lb \nu_{k-1}+1,\nu_{k}\rb=
  \quad e^{2\pi i\nu_{k}}\;\Upsilon_{k-1,k}\lb \nu_{k-1},\nu_{k}\rb,\\
  &\Upsilon_{k-1,k}\lb \nu_{k-1},\nu_{k}+1\rb=
   e^{-2\pi i\nu_{k+1}}\Upsilon_{k-1,k}\lb \nu_{k-1},\nu_{k}\rb,
  \end{align*}
  which may be considered as further evidence for Conjecture~\ref{conjecPI}. 
 From the definition of 
 $\Upsilon_{k-1,k}\lb \boldsymbol\nu\rb$ it is also clear that this quantity should satisfy the cyclic identity
 $\prod_{k=1}^5 \Upsilon_{k-1,k}\lb \boldsymbol\nu\rb=1$. 
 It indeed holds for the above result (\ref{UpsPI}). However, the verification is not completely trivial; the relation
 \ben
 \prod_{k=1}^5\hat G\lb\nu_k\rb=e^{-\pi i/6}\prod_{k=1}^5
 \lb2\pi\rb^{\nu_k}e^{\frac{\pi i \nu_k^2}{2}-2\pi i \nu_{k-1}\nu_k}
 \ebn
 turns out to be equivalent to Abel's five-term identity
 \ben
 \sum_{k=1}^5 L\lb v_k\rb=\frac{\pi^2}{2},
 \ebn
 satisfied by the Rogers $L$-function $L\lb z\rb=\operatorname{Li}_2\lb z\rb+\frac12\ln z\ln\lb1-z\rb$. 
 \end{rmk}


\begin{thebibliography}{10000} 
   \bibitem[BT]{BT} E. L. Basor, C. A. Tracy, {\it Some problems associated with the asymptotics
    of $\tau$-functions},   Surikagaku (Mathematical Sciences) {\bf 30}, no. 3,  (1992), 71--76.       
   
    \bibitem[Ber]{Bertola} M. Bertola, \textit{The dependence on the 
    monodromy data of the
    isomonodromic tau function}, \href{http://dx.doi.org/10.1007/s00220-009-0961-7}{Comm. Math. Phys.~\textbf{294}, (2010),  539--579}; \href{http://arxiv.org/abs/0902.4716}{arXiv: 0902.4716 [nlin.SI]};
    corrig.	\href{http://arxiv.org/abs/1601.04790}{arXiv:1601.04790 [math-ph]}.
    
  \bibitem[BLMST]{BLMST} G. Bonelli, O. Lisovyy, K. Maruyoshi,
  A. Sciarappa, A. Tanzini, {\it On Painlevé/gauge theory correspondence}, \href{http://arxiv.org/abs/1612.06235}{arXiv:1612.06235 [hep-th]}.
  
   \bibitem[Bou]{Boutroux} P. Boutroux, {\it Recherches sur les transcendantes de M. Painlevé et l'étude asymptotique des équations différentielles du sécond ordre}, \href{http://eudml.org/doc/81327}{Ann. Sci. Ecole Norm. Sup. \textbf{30}, (1913), 255--375}; \href{http://eudml.org/doc/81348}{Ann. Sci. Ecole Norm. Sup. \textbf{31}, (1914), 99--159}.
  
  
    \bibitem[DIK]{DIK} 
    P. Deift, A. Its, I. Krasovsky, {\it Asymptotics of the Airy-kernel determinant},
  \href{http://dx.doi.org/10.1007/s00220-007-0409-x}{Comm. Math. Phys. {\bf 278}, (2008), 643--678};  \href{http://arxiv.org/abs/math/0609451}{arXiv:math/0609451 [math.FA]}.  

 \bibitem[DKV]{DKV}
 P. Deift, I. Krasovsky, J. Vasilevska, {\it Asymptotics for a determinant with a confluent hypergeometric kernel},
 \href{http://dx.doi.org/10.1093/imrn/rnq150}{Int. Math. Res. Not. \textbf{2011},  (2011), 2117--2160}; \href{http://arxiv.org/abs/1005.4226}{arXiv:1005.4226 [math-ph]}.        
   
   \bibitem[Del]{Del} E. Delabaere, {\it Divergent Series, Summability and Resurgence III. Resurgent methods and the first Painlevé equation}, \href{http://dx.doi.org/10.1007/978-3-319-29000-3}{Lecture Notes in Math. \textbf{2155}, Springer, (2016)}; \href{http://hal.archives-ouvertes.fr/hal-01067086}{http://hal.archives-ouvertes.fr/hal-01067086}.
   
   
   \bibitem[DGK]{DGK}
   B. Dubrovin, T. Grava, C. Klein, {\it On universality of critical behavior in the focusing nonlinear
   Schr\"o\-din\-ger  equation,  elliptic  umbilic  catastrophe  and  the  tritronquée  solution  to  the  Painlevé-I
   equation}, \href{http://dx.doi.org/10.1007/s00332-008-9025-y}{J.~Nonlin. Sci. \textbf{19}, (2009), 57--94};
   \href{http://arxiv.org/abs/0704.0501}{arXiv:0704.0501 [math.AP]}.
   
   \bibitem[Ehr]{Ehrhardt} 
  T. Ehrhardt, {\it Dyson's constant in the asymptotics of the
  Fredholm determinant of the sine kernel}, \href{http://dx.doi.org/10.1007/s00220-005-1493-4}{Comm. Math. Phys. {\bf 262}, (2006), 317--341}; \href{http://arxiv.org/abs/math/0401205}{arXiv:math/0401205 [math.FA]}.   
   
   \bibitem[FIKN]{FIKN} A. S. Fokas, A. R. Its, A. A. Kapaev, V. Yu. Novokshenov, \textit{Painlev\'e transcendents: the Riemann-Hilbert approach}, Mathematical Surveys and Monographs~\textbf{128}, AMS, Providence, RI, (2006).
   
   \bibitem[HRZ]{HRZ}
   A. N. W. Hone, O. Ragnisco, F. Zullo, 
   {\it Properties of the series solution for Painlevé I},
   \href{http://dx.doi.org/10.1080/14029251.2013.862436}{J. Nonlin. Math. Phys.~\textbf{20}, (2013), 85--100}; \href{http://arxiv.org/abs/1210.6822}{arXiv:1210.6822 [math.CA]}.
   
 \bibitem[ILT13]{ILT13} N. Iorgov, O. Lisovyy, Yu. Tykhyy, \textit{Painlev\'e VI connection problem and monodromy of
 $c = 1$ conformal blocks}, \href{http://dx.doi.org/10.1007/JHEP12(2013)029}{JHEP~\textbf{12}, (2013), 029};
 \href{http://arxiv.org/abs/1308.4092}{arXiv:1308.4092 [hep-th]}.   
   
   
  \bibitem[ILP]{ILP} A. R. Its, O. Lisovyy, A. Prokhorov,
  \textit{Monodromy dependence and connection formulae for isomonodromic tau functions}, (2016), \href{http://arxiv.org/abs/1604.03082}{arXiv:1604.03082 [math-ph]}. 
  
  \bibitem[ILT14]{ILT14} A. R. Its, O. Lisovyy, Yu. Tykhyy,
    \textit{Connection problem for the sine-Gordon/Painlevé III tau function and irregular conformal blocks}, \href{http://dx.doi.org/10.1093/imrn/rnu209}{Int. Math. Res. Not.~\textbf{2015}, (2014), 8903--8924}; \href{http://arxiv.org/abs/1403.1235}{arXiv:1403.1235 [math-ph]}.   
  
  
  \bibitem[IP]{IP} A. Its, A. Prokhorov, \textit{Connection problem for the tau-function of the sine-Gordon reduction of Painlev\'e-III equation via the Riemann-Hilbert approach}, \href{http://dx.doi.org/10.1093/imrn/rnv375}{Int. Math. Res. Not.
  \textbf{2016}, (2016), 6856--6883}; \href{http://arxiv.org/abs/1506.07485}{arXiv:1506.07485 [math-ph]}.  
  
  \bibitem[IS]{IS} K. Iwaki, A. Saenz, {\it Quantum Curve and the First Painlevé Equation}, \href{http://dx.doi.org/10.3842/SIGMA.2016.011}{SIGMA~\textbf{12}, (2016), 011}; \href{http://arxiv.org/abs/1507.06557}{arXiv:1507.06557 [math-ph]}.

 
   \bibitem[JMU]{JMU} M. Jimbo, T. Miwa, K. Ueno,  {\it Monodromy preserving deformation of linear ordinary differential equations with rational coefficients. I}, 
   \href{http://dx.doi.org/10.1016/0167-2789(81)90013-0}{Physica {\bf D2}, (1981), 306--352}.
   
  \bibitem[JK]{JK} N. Joshi, M. Kruskal, {\it An asymptotic approach to the  connection problem for the first and the second Painlevé equations},
  \href{http://dx.doi.org/10.1016/0375-9601(88)90415-X}{Phys. Lett. \textbf{A130}, (1988), 129--137}.
      
   
   \bibitem[Kap1]{Kapaev88}
   A. A. Kapaev, {\it Asymptotic behavior of the solutions of the Painlevé equation of the first kind}, \href{http://www.mathnet.ru/php/archive.phtml?wshow=paper&jrnid=de&paperid=6681&option_lang=rus}{Differentsial'nye
   Uravneniya~{\bf 24}, no. 10, (1988), 1684--1695}; (in Russian).
   
   \bibitem[Kap2]{Kapaev93}
   A. A. Kapaev, {\it Global asymptotics of the first Painlevé transcendent}, preprint INS $\sharp$225, Institute for Nonlinear Studies, Clarkson University, Potsdam, (1993).
   
    \bibitem[Kap3]{Kapaev3}
      A. A. Kapaev, {\it Quasi-linear Stokes phenomenon for the Painlevé first equation}, \href{http://dx.doi.org/10.1088/0305-4470/37/46/005}{J. Phys. \textbf{A37}, (2004),  11149--11167}; \href{http://arxiv.org/abs/nlin/0404026}{arXiv:nlin/0404026 [nlin.SI]}.
   
   \bibitem[KK]{KK} A. A. Kapaev, A. V. Kitaev, {\it Connection formulae for the first Painlevé transcendent in the complex domain}, \href{http://dx.doi.org/10.1007/BF00777371}{Lett. Math. Phys.~\textbf{27}, (1993), 243--252}.    
   
  \bibitem[Mas]{Masoero} 
  D. Masoero, {\it Poles of integrále tritronquée and anharmonic oscillators. A WKB approach}, \href{http://dx.doi.org/10.1088/1751-8113/43/9/095201}{J. Phys.	\textbf{43},
  (2010), 095201}; \href{http://arxiv.org/abs/0909.5537}{arXiv:0909.5537 [math.CA]}.
  
  \bibitem[Tak]{Takei} 
    Y. Takei, {\it On the connection formula for the first Painlevé equation --- from the viewpoint of the exact WKB analysis}, \href{http://repository.kulib.kyoto-u.ac.jp/dspace/handle/2433/59957}{S\={u}rikaisekikenky\={u}sho K\={o}ky\={u}roku~\textbf{931}, (1995), 70--99}.
    
  \bibitem[Tra]{Tracy}  
 C. A. Tracy, {\it Asymptotics of the $\tau$-function arising in the two-dimensional Ising model},
 \href{http://dx.doi.org/10.1007/BF02102065}{Comm. Math. Phys. {\bf 142}, (1991), 297--311}.   
   
   
   \end{thebibliography}
 \end{document}